\documentclass{ws-ijmpd}
\usepackage[super,compress]{cite}
\usepackage{hyperref}
\usepackage{xcolor}
\begin{document}

\markboth{Rinsy Thomas, J Thomas, S P Surendran and M Joy}
{Gravitational wave production after inflation for a hybrid inflationary model}
%{Instructions for Typing Manuscripts (Paper's Title)}

\catchline{}{}{}{}{}
% 
%%%%%%%%%%%%%%%%%%%%%%%%%%%%%%%%%%%%%%%%%%%%%%%%%%%%%%%%%%%%%%%%%%%%

\title{GRAVITATIONAL WAVE PRODUCTION AFTER INFLATION FOR A HYBRID INFLATIONARY MODEL}

\author{RINSY THOMAS$^{1,3}$\footnote{Corresponding author}, JOBIL THOMAS $^{2}$, SUPIN P SURENDRAN$^{3}$ and MINU JOY$^{4}$}

% The "\note" macro will give a warning: "Ignoring empty anchor..."
% you can safely ignore it.
\address{$^{1}$ Department of Physics, CMS College,Kottayam, 686001, India \\rinsy@cmscollege.ac.in\\
$^{2}$ Department of Physics, Indian Institute of Technology Hyderabad, 502285, India \\
$^{3}$School of Pure and Applied Physics, Mahatma Gandhi University, Kottayam,
 686560, India\\
 $^{4}$Department of Physics, Alphonsa College, Pala, 686574, India}

% e-mail addresses: one for each author, in the same order as the authors
%\emailAdd{rinsy@cmscollege.ac.in}

\maketitle

\begin{history}
\received{Day Month Year}
\revised{Day Month Year}
\end{history}
\begin{abstract}
 We discuss a cosmological scenario with a stochastic background of gravitational waves sourced by the tensor perturbation due to a hybrid inflationary model with cubic potential. The tensor-to scalar ratio for the present hybrid inflationary model is obtained as $r \approx 0.0006$. Gravitational wave spectrum of this stochastic background, for large-scale CMB modes, $10^{-4}Mpc^{-1}$ to $1 Mpc^{-1}$ is studied. The present-day energy spectrum of gravitational waves  $\Omega_0^{gw}(f)$ is sensitively related to the tensor power spectrum and r which is, in turn, dependent on the unknown physics of the early cosmos. This uncertainty is characterized by two parameters: $\hat{n_t}(f)$ logarithmic average over the primordial tensor spectral index and $\hat{w}(f)$ logarithmic average over the effective equation of state parameter. Thus, exact constraints in the $\hat{n_t}(f)$-$\hat{w}(f)$ plane can be obtained by comparing theoretical  constraints of our model on r and $\Omega_0^{gw}(f)$. We obtain a limit on $\hat{w}(10^{-15}Hz) < 0.33 $ around the modes probed by CMB scales. 

\end{abstract}


%\ccode{PACS numbers:}

%\tableofcontents

\section{Introduction}	

Primordial gravitational waves (PGW)\cite{Starobinsky:1979ty} have  great potential to explore the physics of the early universe. The stochastic background of gravitational waves (GWs) are produced either during the inflationary period\cite{Rmodel, PhysRevD.23.347, nw} or at the end of inflation but before the Big Bang nucleosynthesis (BBN). Finding such a gravitational wave (GW)\cite{Grishchuk} directly or indirectly provides a unique opportunity to examine inflation.
 
 All relevant scales in cosmology were inside the horizon before inflation began since the comoving Hubble radius was quite large. During the inflationary phase, comoving Hubble radius decreases and the relevant modes left the horizon. The amplitude of fluctuations were almost frozen on super Hubble scales and turned into metric perturbations\cite{Starobinsky:1979ty}. The metric perturbations generated during inflation can be explained by the scalar or curvature perturbations and the tensor perturbations. The scalar fluctuations of the metric that are coupled to the matter density are accountable for the large-scale cosmological structure. The tensor fluctuations in the gravitational metric causes perturbations in the cosmic microwave background (CMB)\cite{ Rubakov-115, Starobinsky:1985ww, Markevich, dodelson:2003}. A key prediction of inflation is the almost scale-invariant nature of scalar\cite{Starobinsky-117, Hawking_1982, PhysRevLett.49.1110} and tensor power spectrum \cite{Starobinsky:1979ty, Easther}. A perfect scale invariance for the scalar power spectrum $(n_s = 1)$ is possible only for very few models \cite{starobinskyInflatonFieldPotential2005}, although a slightly red spectrum ($n_s\lesssim1$) seems to be a prediction  of the simplest plausible one-parameter family of inflationary models, encompassing R + $R^2$ model\cite{Rmodel} and new\cite{nw} and chaotic inflationary model\cite{cha}. A prominent example of a more complicated inflationary model belonging to the slow-roll class is hybrid inflation \cite{lindeHybridInflation1994}. For all observational uses, the tensor spectrum is scale-invariant and the tensor tilt $n_t \simeq 0$. The quantum mechanical fluctuations of spacetime during inflation provide an almost scale-invariant spectrum of GWs in the present period and are the best-known probable source for the gravitational wave background(GWB). Another phenomenon that is most effective at generating GWs in the early Universe is the parametric resonance due to preheating after inflation \cite{PhysRevD.56.653, Easther}.

The physics of inflation can be significantly constrained by the Planck and Keck/BICEP2 experiments. Since GW is accountable  for the B-mode polarization of CMB anisotropies, its spectrum should in theory be measurable by CMB polarisation experiments.  If slow-roll inflation by a single field is confirmed, tensor to scalar power spectrum ratio (r),  can be used to directly determine the inflationary energy scale \cite{PhysRevD.49.739}. In conjunction with the limitations of the scalar spectral index, the combination of Planck 2018, BICEP2 and Keck array data slightly limits the tensor to scalar ratio at comoving wavenumber $k=0.002Mpc^{-1}$ as $r_{0.002}= 0.06 $\cite{aghanimPlanck2018Results2020} and is effective in identifying inflationary models. For example, Planck 2018 data, strongly disagrees with the monomial potentials like $V(\phi)=\lambda M_{pl}^4\left(\phi/M_{pl}\right)^p$ having $p\geq2$ while the models with p=1 and p=2/3 are more compatible with recent Planck results. The Planck 2018 data shows that a quartic potential gives a more accurate match than a quadratic one within the class of hilltop inflationary models\cite{Boubekeur_2005}. For a considerable percentage of their parameter space, D-brane inflationary models\cite{Shamit} fit Planck and BK15 data well. Planck 2018 and BK15 results\cite{planckcollaborationPlanck2018Results2020} severely disfavors the hybrid model of inflation inspired by logarithmic quantum corrections in spontaneously broken supersymmetric theories (SB Susy)\cite{PhysRevLett.73.1886}. This disagreement has important implications for the viability of the SB Susy hybrid inflation model. However there are several hybrid inflationary models such as smooth hybrid inflation\cite{Ahmed}, hilltop hybrid inflation\cite{Boubekeur_2005} that are in agreement with Planck's data.

 PGW could also be produced by a variety of post-inflationary mechanisms. Because of the wide range of processes, it is significant to place tight constraints on the spectral energy density of gravitational waves, $\Omega_{0}^{gw}(f)$, for the whole range of detectable frequencies. Constraints may be established using data obtained from several sources, including the CMB, gravitational wave interferometers, pulsar timing, and the BBN. Measurements of the CMB's temperature and polarisation power spectrum may place limits on $\Omega_{0}^{gw}(f)$. The primordial spectrum of tensor perturbations in the single-field slow-roll inflationary model uses a power law, with parameters defined by the tensor spectral tilt $n_t$  and the tensor-to-scalar ratio r, and it is directly related to $\Omega_{0}^{gw} (f)$. With the assumption that the power-law holds true across several decades in scale, the expected signal may be extrapolated to higher frequencies \cite{clarkeConstraintsPrimordialGravitational2020}. The limits on $\Omega_{0}^{gw}(f)$ can be used to constrain the effective equation of state $\hat{w(f)}$. One of the parameters that determine the shape of the gravitational wave spectrum is $\hat{w}(f)$.

However, there are certain models, such as hybrid inflation, that have a rapid end to inflation and nearly instantaneous conversion to radiation. Hybrid inflation models with symmetry-breaking field connected to a flat direction is seen in several theories of particle physics and can be found in a variety of extensions of the Standard Model, including string theory and supersymmetric theories \cite{lindeHybridInflation1994}. The fundamental advantage of hybrid models is that the scale of inflation can range from grand unified theory (GUT) sizes to TeV scales, whereas most chaotic inflation models can only exist on high energy scales, with Planck scale inflaton values of $V^{1/4}_{inf}$ $10^{16}$ GeV \cite{garcia-bellidoStochasticBackgroundGravitational2007}. Because of the wide range of energy scales present in hybrid inflation, from the GUT scales to the TeV scale, future GW detectors should be able to detect the stochastic background of GWs produced during preheating. In hybrid inflationary models, a peak in the primordial power spectrum\cite{Choi_2021} of the curvature fluctuation on tiny scales leads to the development of a primordial black hole, which may serve as the universe's dark matter component. 

   We discuss a model with two interacting scalar fields in which inflation can be induced  by the slow rolling of the field $\phi$ and ended by a quick rolling of an auxiliary scalar field $\psi$. The inflationary potential is subjected to a rapid minor change in its second derivative (the effective mass of the inflaton), which is the significant feature of this model. A sudden small change in the effective mass of the inflaton causes a slight  change in the primordial spectral index $n_s$, which is followed by decreasing-amplitude oscillations. Such inflationary potential behaviour is the result of a quick phase change experienced by a second scalar field $\psi$ weakly connected to the inflation field, based on theoretical models. This model  is identical to that used to halt inflation in the hybrid inflationary scenario\cite{lindeHybridInflation1994, joyNewUniversalLocal2008}. Tensor to scalar perturbation ratio (r) and scalar spectral index ($n_s$), which characterizes the degree of the scale dependence\cite{Minu Joy_2011, Mukherjee_2015 } of the fluctuation amplitude, are the two most essential observable components of the primordial power spectrum. 
   
   In Sec.\ref{sec2} of this article, we discuss a hybrid inflation model with a step-like discontinuity in its effective mass. In Sec.\ref{sec3}  we study the primordial gravitational wave produced by the present hybrid inflationary model and also obtain the theoretical constraints on the parameters $r$, $n_t$ of our model that governs the evolution of the early universe, in light of the Planck data. In Sec.\ref{sec4} we discuss the average tensor tilt $\hat{n_t}(f)$ and the effective equation of state $\hat{w}(f)$. Conclusions and discussions are covered in Sec.\ref{sec5} .

\section{Hybrid Inflationary Model}
\label{sec2}

 We consider the hybrid inflationary model with cubic potential. In this inflationary model, the evolution of effective mass features a step-like discontinuity.
\begin{align} \label{eq:1}
{
 V(\psi,\phi)\quad=\quad \frac{M^4\lambda}{4}(1+c\phi^3)\,-\,\frac{1}{2}\lambda\psi^2M^2\,+\,\frac{1}{4}\lambda\psi^4\,+\,\frac{1}{2}\lambda'\psi^2\phi^2
 }
\end{align}
\hspace{1cm} with $c>0$.

The parameter $\lambda$ determines the strength of the self-coupling of the field $\psi$
and $\lambda'$ determines the mutual interaction between $\psi$ and $\phi$. The slowly rolling inflation field $\phi$ does not account for the majority of the energy density in hybrid inflation. An auxiliary field $\psi$ takes this role, which is held in place by its interaction with $\phi$ until $\phi$ falls below a critical value $\phi_c$. When this occurs, $\psi$ has been destabilized and inflation comes to an end by rolling towards its true vacuum \cite{liddleCosmologicalInflationLargeScale2000, MinuJoy_2009}.
We can deduce from (\ref{eq:1}) that the effective mass of the field near $\psi\,=\,0$ is given by
 \begin{align}
     m_\psi^2=\frac{d^2V}{d\psi^2}=-\lambda M^2+\lambda'\phi^2
 \end{align}
in order that $m_\psi^2>0$ if $\phi>\phi_c$ and $  m_\psi^2<0$ if $\phi<\phi_c$, where  $\phi_c=M\sqrt{\frac{\lambda}{\lambda'}} $ is the field's critical value, at which the curvature of the potential $ V (\psi,\phi) $ vanishes along the $\psi$ direction. $m_\psi^2>0$ makes certain that field rolls down from large $\phi$ along the $\psi=0$ channel, until it reaches the point of instability.  As the $\psi$=0 solution becomes unstable and results in a rapid cascade (waterfall) of $\psi$ towards the minimum of its potential \cite{Gong_2022}. At late times, $m_\psi^2<0$, field rolls to their true minimum at $\phi=0$, $\psi=\mp M$. Inflation's end corresponds to a second-order phase transition with no barrier. As the weak second order phase transition approaches $\phi >M\sqrt{\lambda/\lambda'}$ and $\psi=0$, as a result

\begin{align}
    V(\phi)\quad=\quad\frac{M^4\lambda}{4}(1+c\:\phi^3)
\end{align}

At the point of phase transition, $\phi = \phi_c $ and 
\begin{align}
    V(\phi_c)\quad=\quad\frac{M^4\lambda}{4}(1+\kappa)
\end{align}
 where $\kappa= c\:M^3\left( \frac{\lambda}{\lambda'}\right)^{3/2}$.  
If $\kappa>1$, the correction to the vacuum energy density $V(0,0)=\frac{M^4\lambda}{4}$ due to $c\phi^3$ term is substantial, and if $\kappa<1$, then the vacuum energy density is unaffected. It is clear that the slow roll condition is valid before the phase transition and the slow roll condition $\frac{\left| V''\right|}{H^2}<<1$ implies,
\begin{align}\label{eq:2}
    M\ll\frac{(1+\kappa)}{18\:c \:M_{pl}^2}\sqrt{\frac{\lambda'}{\lambda}}
\end{align}
where $M_{pl}$ is the reduced planck mass. 

Immediately following the phase transition at $\phi_c $, $\phi\:<\: M\:\sqrt{\frac{\lambda}{\lambda'}}$,\quad $\psi^2\:=\:\frac{\lambda\:M^2\:-\:\lambda'\phi^2}{\lambda}$

\begin{align}
 V(\phi)=\frac{M^4\lambda}{4}\left(1+c\:\phi^3\right)\:-\: \frac{\left( \lambda M^2-\lambda'\phi^2\right)^2}{4\:\lambda}   
\end{align}
The condition that slow-roll continues to be valid immediately after the phase transition provides
\begin{align}\label{eq:5}
    M^2\:\ll\:\frac{1}{\lambda'M_{pl}^2}
\end{align}
The product of (\ref{eq:2}) and (\ref{eq:5}) yields the constraint below.
 \begin{align}\label{eq:3}
     M^3\:\ll\:\frac{(1+\kappa)}{18\:c\:Mpl^4\sqrt{\lambda\:\lambda'}}
 \end{align}
The motion of $\psi$ is rapid in contrast to the field $\phi$, where the field $\phi$ slowly rolls down the potential. The condition $\dfrac{\left|\frac{\partial^2 V}{\partial \psi^2}\right|}{H^2}\:\gg\:1$ is valid through all times, with an exception of a very short interval $\Delta t \ll  H^{-1}$ around the transition if
\begin{align}\label{eq:4}
    M\:\gg\:\frac{72\:c\:M_{pl}^4}{(1+\kappa)^2}\sqrt{\frac{\lambda}{\lambda'}}
\end{align}
By applying the slow roll condition before and after the transition it is simply observed from (\ref{eq:2}) and (\ref{eq:4}) that $\lambda' \ll \lambda$ in our model. The $\psi$ field's self-coupling should be substantially greater than its coupling to the inflation field $\phi$. The constraint given by (\ref{eq:3}) also satisfies this.

The precise value of the number of e-folds (N) in inflationary cosmology is a free parameter. As a generous calculation, we take N=60 which is uniquely defined by the value of $\kappa, \, \phi_c$ and $\phi_e$ where $\phi_{e}$ is the value of the inflaton field where the inflation ends. The inflationary curvature fluctuation on large scales can be compared to the observed CMB fluctuation to infer the values of parameters ($\lambda\:,\:\lambda'\:,\:M$) in our model. It's possible to approximate the perturbation spectrum as
\begin{align}\label{eq:6}
    P_R(k)\:=\:\frac{1}{12\pi^2 M_{pl}^6}\frac{V^3}{V'^2}\:=\:\frac{1}{432\pi^2 M_{pl}^6}\frac{\lambda'^2}{\lambda \,c^2}\left(1+\kappa^3\right)^3
\end{align}
(\ref{eq:6}) is connected to the dimensionless density contrast $\delta_H(k)$ as,
\begin{align}
    \delta^2_H(k)\:=\:\frac{4}{25}\left({\frac{g(\Omega_m)}{\Omega_m}}\right)^2P_R(k)
\end{align}
where, {\cite{liddleCosmologicalInflationLargeScale2000}\\
\begin{align}
    g(\Omega_m)=\frac{5\,\Omega_m}{2}\left(\frac{1}{70}\,+\,\frac{209\Omega_m\,-\,\Omega_m^2}{140}\,+\,\Omega_m^{\frac{4}{7}}\right)^{-1}.
\end{align}

For an LCDM Universe, the four-year COBE data suggests 
\cite{liddleCosmologicalInflationLargeScale2000, bunnFouryearCOBENormalization1996}
\begin{align}
    \delta_h\:=\:1.91\,\times \,10^{-5}\frac{exp\left[1.01\left(1-n\right)\right]}{\sqrt{1+f\left(\Omega_m\right)r}}\Omega_m^{-0.8-0.05\log \Omega_m}\,\times\,\left[1-0.18(1-n)\Omega_\lambda-0.03r\Omega_\lambda\right],
\end{align}
where $f(\Omega_m)\:=\:0.75\,-\,0.13\Omega^2_\lambda$ and the tensor to scalar ratio, indicated by $r\simeq16\epsilon_0$.
Taking $\lambda=0.1$ and assuming a spatially flat LCDM universe with $\Omega_m=0.3153$ and $\Omega_\lambda$=0.6847 and using the relations $\kappa\,=\,c\phi_c^3$ and $\phi_c=M\sqrt{\frac{\lambda}{\lambda'}}$  parameter value c, $\lambda'$ and M can be determined.\\

\begin{table}[ph]
\tbl{Typical parameter values for the hybrid inflationary potential.}
{\begin{tabular}{@{}cccc@{}} \toprule
$\kappa$ & $c M_{pl}^3$ & $\lambda'$ & $M/M_{pl}$\\
%& (Rad/s) & (Rad/s) \\ \colrule
0.18\hphantom{00} & \hphantom{0}0.0000307779 & \hphantom{0}$2.23523\times10^{-8}$ & 0.008518 \\ \botrule
\end{tabular} \label{ta0}}
\end{table}

As given in Table~\ref{ta0},  we conclude that the value of M for this model is in the GUT range.  All inequalities (\ref{eq:2}), (\ref{eq:5}), and (\ref{eq:4}) are satisfied for these parameter values.
\section{Primordial Gravitational wave spectrum}
\label{sec3}
Traces of  primordial gravitational waves in the low-frequency range can be obtained by CMB polarisation observations \cite{PhysRevD.96.063508}. Here we compute the primordial spectra for scalar and tensor perturbations for the hybrid inflationary model. The scalar spectral index is an important parameter in cosmology that characterizes the spectrum of primordial density fluctuations produced during inflation. The scalar power spectrum of this model is obtained as almost scale-invariant with a step in its spectral index $n_s$ at $\phi_c$\cite{MinuJoy_2009} due to the rapid phase transition experienced by $\psi$ that is weakly coupled to inflaton field $\phi$. Around the transition region, there are small oscillations that diminish in both $P_R(k)$ and $n_s(k)$ as we move away from the point of transition %At the point of transition, the magnitude of the running of the spectral index is quite significant, approximately equal to (ns-1), but it also diminishes as it moves away from that point and it is %
as shown in Fig. \ref{f1}. The scale dependence of the present hybrid model's spectrum is dominated by a red tilt with the spectral index, $n_s=0.962$ which is in agreement with Planck 2018 data\cite{planckcollaborationPlanck2018Results2020}.

Tensor primordial power spectrum is obtained as scale-invariant for the present hybrid model. The primordial scalar and tensor power spectrum play a crucial role in determining the tensor-to-scalar ratio, a key parameter in understanding the early universe. We observed a small anomaly in the tensor-to-scalar ratio at the transition point, which arises from the localized feature of the scalar power spectrum. As it is clear from Fig. \ref{fig:ratio} an upper limit on the tensor to scalar ratio is obtained as $r \approx 0.0006$ which is in agreement with the recent Planck results \cite{aghanimPlanck2018Results2020}. When integrating Planck 2018, BICEP2/Keck, and Baryon Acoustic Oscillations (BAO) data, the best constraint on the tensor-to-scalar ratio at 95\% confidence level is $r_{0.002}<0.056$\cite{aghanimPlanck2018Results2020, keckarrayandbicep2collaborationsConstraintsPrimordialGravitational2018}.
\begin{figure}[!h]
\centerline{\psfig{file=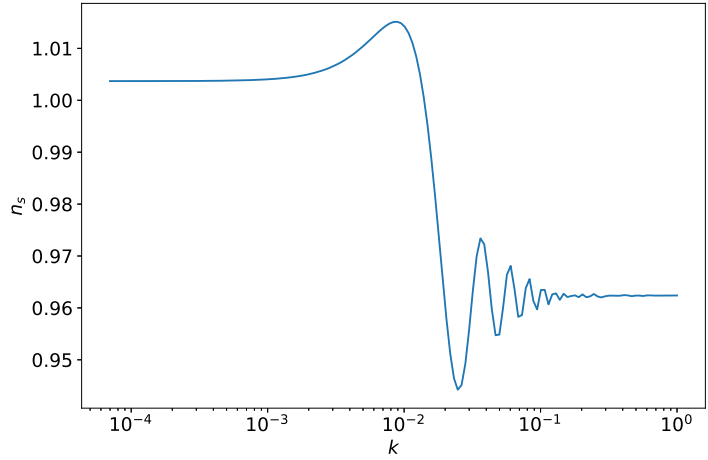,width=10.7cm}}
\vspace*{10pt}
\caption{The primordial spectral index, $n_s$ versus the wavenumber, k for the hybrid inflationary model \label{f1}}
\end{figure}
%Overall, our findings suggest that the hybrid inflationary model is a viable cosmological model that is consistent with Planck data.
\begin{figure}[!h]
\centerline{\psfig{file=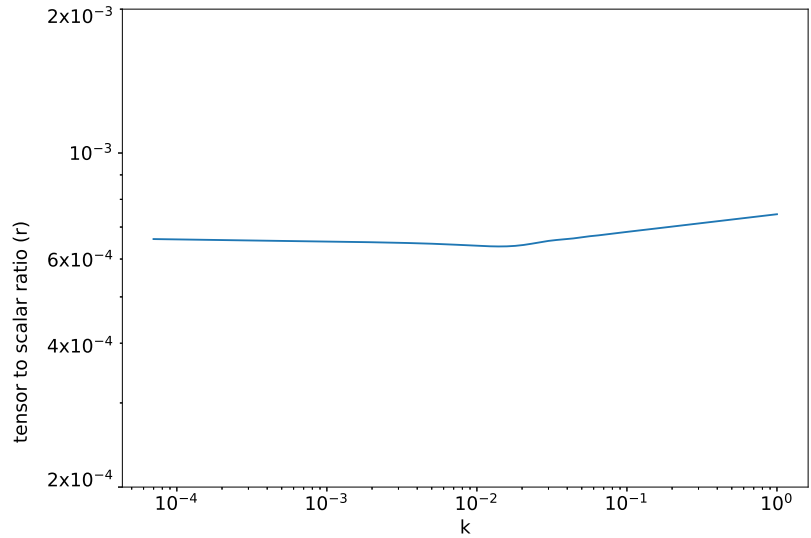,width=10.7cm}}
\vspace*{10pt}
\caption{The tensor-to-scalar ratio, r versus the wavenumber, k for the hybrid inflationary model \label{fig:ratio}}
\end{figure}
The tensor-to-scalar ratio, r is a measure of the energy scale of inflation, 
$V^{\frac{1}{4}}=(100r)^{\frac{1}{4}}10^{16}$ GeV. 
The red-tilted spectral index can only be very small, $-n_t\le 0.00066\ll1$, in the slow roll scenario. Therefore, the tensor spectrum is extremely close to being scale-invariant, at least at scales close to the CMB .

GWs are characterized by the tensor perturbations $h_{ij}$, which are gauge-invariant and transverse-traceless in nature. In Fourier space, the equation of motion for $h_{ij}$ can be written in the form of a wave equation. This equation can be obtained from the linearized Einstien field equations. In terms of conformal time derivatives\cite{ PhysRevD.73.123515}.
\begin{align}
h^{''}_{ij}(\tau,k)+\left(\frac{2a'}{a}\right) h^{'}_{ij} (\tau,k)+k^2 h_{ij}(\tau,k)\quad=\quad 16 \pi G a^2 \Pi_{ij}(\tau,k)
\end{align}
where $\Pi_{ij}(\tau,k)$ is the anisotropic part of the stress tensor. The time evolution of $h_{ij}(\tau,k)$ can be expressed in terms of the primordial gravitational wave modes, denoted $h_{ij}^{prim}$, refers to the specific pattern of gravitational fluctuations that emerged from the horizon during the inflationary period and the transfer function $T(\tau,k)$ which characterises how gravitational wave modes evolve within the horizon after they have entered it\cite{PhysRevD.73.123515}.
\begin{align}
    h_{ij}(\tau,k)=h_{ij}^{prim} T(\tau,k)
\end{align}
The tensor power spectrum at any time is related to primordial tensor power spectrum $P_T(k)$ as,
\begin{align}
P_T(\tau,k)=P_{T}(k)\left[T(\tau,k)\right]^2
\end{align}
The gravitational wave energy density parameter today can be written as\cite{PhysRevD.73.123515,caprini},
\begin{align}
    \Omega^{gw}_{0}(k)\equiv\frac{P_T(k)}{12a_0^2 H_0^2}\left[T^{'}(k,\tau_0)\right]^2
\end{align}

 The primordial tensor spectrum created during inflation, as well as, more importantly, the rate of the universe's expansion from the end of inflation to the present, determine the fractional energy density in the primordial gravitational waves that have been observed today \cite{Chowdhury_2022}.For an expanding universe, which includes non-relativistic matter and radiation, the present-day gravitational wave spectrum $\Omega_{0}^{gw}(k)$ can be defined in terms of wave number k and the tensor power spectrum $P_T(k)$ as\cite{laskyGravitationalWaveCosmology292016, PhysRevD.48.4613, PhysRevD.73.123515}, 
\begin{align}\label{Eq:GW}  
    \Omega_{0}^{gw}(k)h^2=\frac{3}{128}\Omega_rh^2P_T(k)\left[\frac{1}{2}\left(\frac{k_{eq}}{k}\right)^2+\frac{16}{9}\right]
\end{align}
where the uncertainty in the Hubble parameter is given by $h=0.674$, $H_0=100hkms^{-1}Mpc^{-1}$, $\, k_{eq}=\frac{\sqrt{2}H_0\Omega_m}{\sqrt{\Omega_r}}$ is the wavenumber of a mode that enters the horizon at matter radiation equality, $\Omega_r$ is the density of relativistic species and $\Omega_m$ corresponds to matter density.

The gravitational wave energy spectrum as a function of frequency, f and wave number, k for the present hybrid inflationary model obtained using Eq.(\ref{Eq:GW}) is given in Fig. \ref{fig:GW}, which is consistent with recent Planck and BICEP2/Keck results\cite{planckcollaborationPlanck2018Results2020, clarkeConstraintsPrimordialGravitational2020}. The energy density spectrum of inflationary gravitational waves observed today\cite{Figueroa_2019}, which exhibits (quasi-)scale invariance, represents the amplitude for the modes that passed the horizon during the radiation-dominated era. The range of $\Omega_{0}^{gw}(k)h^2$ for frequencies $10^{-19}$ Hz to $10^{-15}$Hz is obtained as $10^{-14}$ to $10^{-18} $ respectively. Because GWs are formed causally within the Hubble volume at the time, and simple inflationary models often occur at the GUT scale, the typical wave length of these GWs is much shorter than LIGO scales. The scope of existing and upcoming interferometers and PTA experiments is extensive, distinguished by their ability to investigate the gravitational wave spectrum at various frequencies, complementing one another.  PTAs are expected to examine gravitational waves within the range of approximately $10^{-9}$ to $10^{-7}$ Hz. Furthermore, it is anticipated that the upcoming interferometers and PTA experiments will be capable of detecting gravitational waves in an even lower frequency range than previously mentioned, from $10^{-19}$ to $10^{-17}$ Hz, providing a more comprehensive understanding of the gravitational wave spectrum.
\begin{figure}[!hpb]
\centerline{\psfig{file=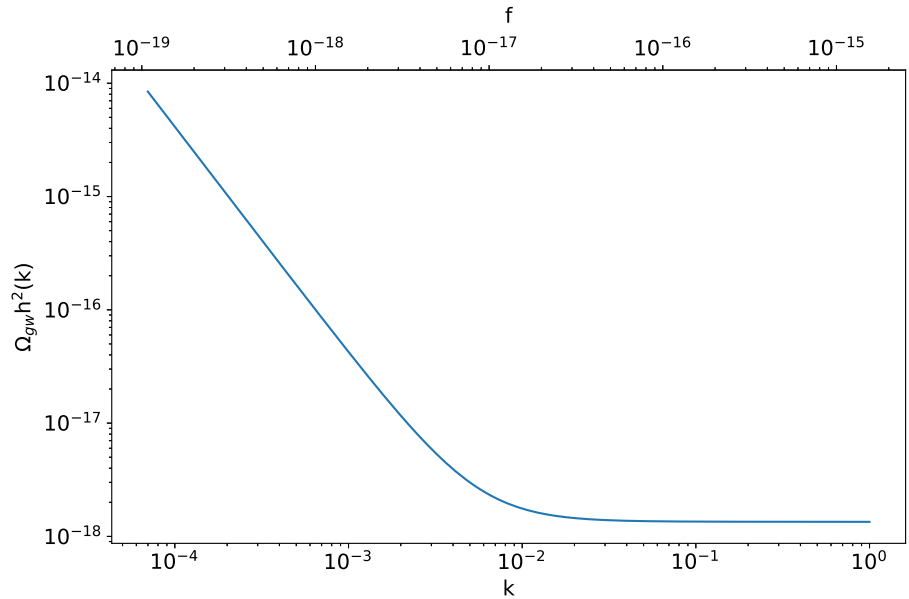,width=12cm}}
\vspace*{10pt}
\caption{The gravitational wave energy spectrum as a function of frequency,f and wave number,k for the hybrid inflationary model\label{fig:GW}}
\end{figure}
\section{Constraining the evolution of early universe}
\label{sec4}
 %The factors that control the evolution of the universe are related to the gravitational-wave spectrum, $\Omega_0^{gw}(f)$.
 Primordial gravitational wave measurements probe two fundamental quantities, the tensor to scalar ratio $r$ and $\Omega_0^{gw}(f)$. The tensor-to-scalar ratio r is constrained on long wavelengths by CMB polarisation measurements and the present-day primordial  gravitational-wave energy spectrum, $\Omega_0^{gw}(f)$ is constrained on shorter wavelengths by numerous techniques. 

 Our analysis is relevant to the primordial gravitational wave spectrum produced during the inflationary epoch and it does not encompass certain scenarios, such as the gravitational-wave spectrum generated by the hypothetical period of preheating following inflation, or the collisions of bubbles after a cosmological phase transition. These particular production mechanisms give rise to gravitational waves that possess wavelengths shorter than the instantaneous Hubble length at the moment of their generation. 
 
 The relation between $\Omega_0^{gw}(f)$ and r are sensitive to  the logarithmic average of the effective equation of state parameter in the early universe after horizon re-entry ($\hat{w}(f)$) and the logarithmic average of the primordial spectral tilt ($\hat{n_t}(f)$). The equation that relates $\Omega_0^{gw}(f)$ and r is\cite{PhysRevD.78.043531}
\begin{align}\label{eq:7}
    \Omega_0^{gw}(f)=A_1A_2^{\hat{\alpha}(f)}A_3^{\hat{n_t}(f)}r
\end{align}
The exponent $\hat{\alpha}(f)$ given in (\ref{eq:7}) can be defined as,
\begin{align}
    \hat{\alpha}(f) \equiv \left(\frac{3\hat{w}(f)-1}{3\hat{w}(f)+1}\right)
\end{align}
The three factors ${A_1,A_2,A_3}$ appearing in (\ref{eq:7}) are determined  by 

\begin{align}
    A_1=\frac{C_2(k)C_3(k)P_T(k_{cmb})\gamma}{24}
\end{align}
\begin{align}
    A_2=\left(\frac{2\pi f}{H_0}\right)\frac{1}{(1+z_c)\gamma^{\frac{1}{2}}}
\end{align}
\begin{align}
    A_3=\frac{2\pi f}{H_0}\frac{H_0}{\frac{k_{cmb}}{a_0}}
\end{align}
with, 
\begin{align}
    \gamma \equiv \frac{\Omega_{m}g_*(z_c)g_{*s}^{\frac{4}{3}}(z_{eq})}{1+z_{eq} g_*(z_{eq})g_{*s}^{\frac{4}{3}}(z_c)}
\end{align}
where $\Omega_{m} = 0.3153 \pm 0.0073$ is the ratio of present-day nonrelativistic density of matter to the present-day critical density, $H_0=67.4km/s/MPc \pm 0.5$. In particular, $z_{eq}=3387$ denotes the redshift of matter-radiation equality, and $z_c=5.9 \times 10^9$ denotes the maximum redshift at which the universe was largely dominated by radiation. The quantities $g_*(z)$ and $g_{*s}(z)$, determine the universe's total number of degrees of freedom of relativistic particles at redshift z. Based on the currently available observational knowledge of the early cosmos $z_c$ is chosen as the redshift of BBN, $z_{BBN}$. The standard values for $g_*(z_{eq})$and $g_{*s}(z_{eq})$ at matter-radiation equality are $3.36$ and $3.91$; and standard values for $g_*(z_{BBN})$ = $g_{*s}(z_{BBN})$ are $10.75$ at BBN (when the temperature is T = 1 MeV) respectively \cite{Figueroa_2019}. The relationship between comoving wavenumber $k$ and the physical frequency $f$ is given by the relation $\frac{k}{a_0}=2\pi f$ and $a_0$ is the cosmic scale factor for the current universe and conventionally set to 1. The factor $C_2(k)$ bears the behaviour of the background equation of state during the horizon re-entry phase and is known as the horizon crossing factor \cite{PhysRevD.78.043531}.
\begin{align}
    C_2(k)=\frac{\Gamma^2 (\alpha_k+1/2) }{\pi}\left[\frac{2}{\alpha_k}\right]
\end{align}
$\Gamma(x)$ is the gamma function and $\alpha_k \equiv \frac{2}{1+3\Tilde{w_k}}$, $\Tilde{w(k)}$ is the effective equation of state parameter at redshift $z_k$ and its radiation like, $\tilde{w}=1/3$ and $C_2(k)=1$. The tensor anisotropic stress $\pi_{ij}$ (for instance, from free-streaming relativistic particles) in the early universe modifies the primordial gravitational wave signal, and this is captured by the factor $C_3(k)$,
\begin{align}
    C_3(k)=A^2(k)
\end{align}
where $A(k)=\frac{15(14406\Omega_{fs}^4-55770\Omega_{fs}^3+3152975\Omega_{fs}^2-48118000\Omega_{fs}+324135000)}{343(15+4\Omega_{fs})(50+4\Omega_{fs})(105+4\Omega_{fs})(180+4\Omega_{fs})}$, where $\Omega_{fs}$ is the percentage of the critical density that has been relativistically free-streaming at redshift $z_k$. Additionally, $\Omega_0^{gw}(f)$ is damped by free-streaming neutrinos in the range of $10^{-16}$Hz to $10^{-10}$ Hz \cite{campetiMeasuringSpectrumPrimordial2020}. In particular, the damping factor is 0.80313 if we substitute $\Omega_{fs} = 0.4052$, which corresponds to three standard neutrino species.

\begin{figure}[!h]
\centerline{\psfig{file=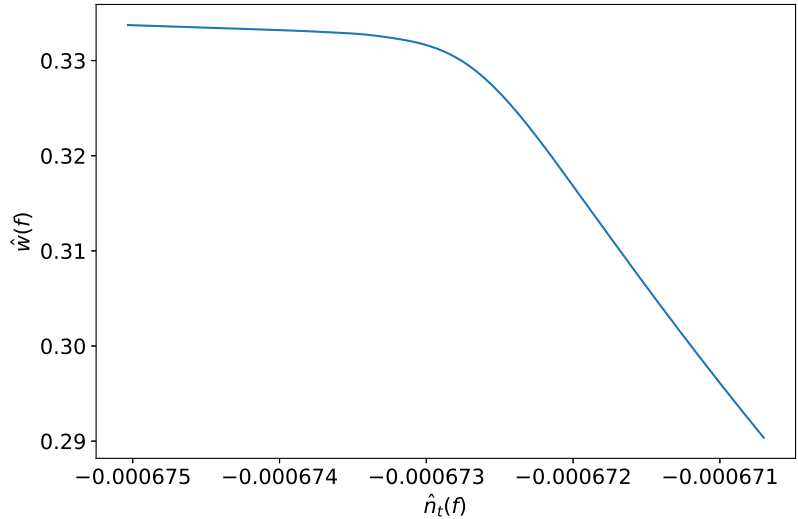,width=12.0cm}}
\vspace*{10pt}
\caption{$\hat{n_t}-\hat{w}$ plane for $r=0.0006$.\label{f4}}
\end{figure}
Since the temperature is rather low (equivalent to energies in atomic physics) when the modes probed by the CMB re-enter the horizon, we are reasonably convinced that neutrinos are the only free-streaming relativistic particles that decouple around the time of BBN.
Hence, $\Omega_0^{gw}(f)$ and r can be used to place constraints in the $\hat{n_t}$-$\hat{w}$ plane and it is depicted in Fig. \ref{f4}. From the plot its clear that the $\hat{w}(f)$ evolve towards $\frac{1}{3}$, which corresponds to a radiation dominated universe. This is due to the fact that at the end of inflation, the inflaton field undergoes a phase transition, triggering the reheating process. In the present hybrid inflationary model also the phase transition and the subsequent reheating  determine the evolution of the equation of state. The presence of non-thermalized fields or interactions that are not instantaneously in thermal equilibrium results in a transient period where $\hat{w}(f)$ has values slightly less than $\frac{1}{3}$\cite{Podolsky2005EquationOS }. During the reheating phase, which occurs before the system reaches thermal equilibrium, $\hat{w}(f)$ increases to a value of around $0.2 - 0.3$. However, as the reheating process progresses, energy transfer and particle production become more efficient, leading to thermalization. This change in $\hat{w(f)}$ happens quickly, within a few e-folds. Eventually, the equation of state reaches the radiation-dominated regime, with an effective equation of state parameter of $\hat{w}(f) = \frac{1}{3}$ \cite{PhysRevLetDai, PhysRevDJul}. Present hybrid inflationary model with $r=0.0006$ and $n_t(10^{-15}Hz) = -0.00066 $ is implying a bound on the effective equation of state parameter $\hat{w}({10^{-15}}Hz) < 0.33$.

\section{Conclusion}
\label{sec5}
A stochastic gravitational wave background will be produced by gravitational field quantum fluctuations according to the inflationary model of the early universe. A unique probe of the fundamental physics of the very early universe is encoded in such radiation. %It is interesting to note that primordial inflationary and reheating phases introduce peculiar contributions and characteristics to the primordial gravitational wave power spectrum. 
A primordial gravitational wave (PGW) background with a broad frequency range and a nearly scale-invariant profile is predicted by cosmological inflation, but its amplitude is too weak to be picked up by upcoming gravitational wave experiments \cite{Chowdhury_2022}. In the present hybrid inflationary model with cubic potential, the inflaton's effective mass changes quickly. To be specific a sudden change in $m^2_{eff}$  satisfying the condition $|m^2_{eff}| \ll H^2$ causes a step in the value of the primordial scalar spectral index $n_s$. It represents a rapid second-order phase change that takes place during inflation, for an auxiliary    scalar field weakly coupled to the inflaton. The primordial tensor power spectrum for such a model is obtained as scale independent. It is also found that this hybrid inflationary model leads to a slightly red-tilted tensor spectrum.  

In conclusion, the present hybrid inflationary model predicts the existence of a stochastic background of gravitational waves. Energy spectrum of PGWs for this inflationary model is plotted in Figure \ref{fig:GW} which is consistent with the gravitational wave spectrum for a slow-roll scenario with Planck and BICEP2/Keck upper limit of $r_{0.002}$ = 0.056 and $n_t$= 0.007\cite{planckcollaborationPlanck2018Results2020, clarkeConstraintsPrimordialGravitational2020}. The relative amplitude of the gravitational wave obtained is also in agreement with the recent BICEP2/Keck and Planck data.  It is interesting to note that for the hybrid inflationary model with $r=0.0006$ and $n_t=-0.00066$ around the modes probed by CMB scales put a strong constraint on the effective equation of state parameter, $\hat{w}(10^{-15}Hz)<0.33$ at the end of inflation.  In particular, we see that at the end of inflation, as the reheating process continues, energy transfer and particle production become more efficient, leading to thermalization and the value $\hat{w(f)}$ evolve towards the equilibrium radiation-dominated value $\frac{1}{3}$ in cases such as those outlined in\cite{PhysRevLett.73.3195, Podolsky2005EquationOS}. For a high energy scale inflationary model expected gravitational wave background emerges at very high frequencies. Eventhough it has large amplitude, this frequency is well above the sensitivity of current experiments, where no detector has yet proven to be sensitive. In the present hybrid inflationary model, inflation occurs at low energy scales. Low energy scale hybrid inflationary models have the benefit of producing background radiation\cite{garcia-bellidoGravitationalWaveBackground2008} that can be detected by upcoming gravitational wave detectors like BBO. PGWs also have an impact on small-scale CMB anisotropies. The characterization and detection of such gravitational wave background radiation which originates from the largely unexplored phases of the universe may provide a very new look during the early cosmos.

\section*{Acknowledgments}

MJ acknowledges the Associateship of IUCAA. We thank the reviewers for the valuable suggestions and comments.

\bibliographystyle{ws-ijmpd}

\begin{thebibliography}{10}

\bibitem{Starobinsky:1979ty}
A.~A.Starobinsky, {\em JETP Lett.} {\bf 30} (1979)   682.

\bibitem{Rmodel}
A.~Starobinsky, {\em Physics Letters B} {\bf 91}  (1980)  99.

\bibitem{PhysRevD.23.347}
A.~H.Guth, {\em Phys. Rev. D } {\bf 23} (1981) 347

\bibitem{nw}
A.~Linde, {\em Physics Letters B} {\bf 108}  (1982)   389.

\bibitem{Grishchuk}
L.~P. Grishchuk, {\em Sov. Phys. JETP} {\bf 40}  (1975)   409.


\bibitem{Rubakov-115}
V.~A. Rubakov, M.~V. Sazhin, A.~V. Veryaskin,{\em Physics Letters B} {\bf 115} (1982) 189.

\bibitem{Starobinsky:1985ww}
A.~A. Starobinsky, {\em Sov. Astron. Lett.} {\bf 11} (1985) 133.

\bibitem{Markevich}
A.~V. Markevich, A.~A. Starobinsky,  {\em Astronomy Letters } {\bf 22} (1996) 431

\bibitem{dodelson:2003}
S.~Dodelson, {\em {Modern Cosmology}} (Academic Press, Elsevier Science, 2003).

\bibitem{Hawking_1982}
S.~W. Hawking, {\em Physics Letters B} {\bf 115} (1982) 295.


\bibitem{Starobinsky-117}
A.~A.Starobinsky, {\em Physics Letters B} {\bf 117} (1982) 175.


\bibitem{PhysRevLett.49.1110}
A.~H. Guth and S.-Y. Pi, {\em Phys. Rev. Lett.} {\bf 49} (1982) 1110.

\bibitem{Easther}
R.~Easther, J.~T. Giblin and E.~A. Lim, {\em Phys. Rev. D} {\bf 77} (2008) 103519.

\bibitem{starobinskyInflatonFieldPotential2005}
A.~A. Starobinsky, {\em Journal of Experimental and Theoretical Physics Letters} {\bf 82}  (2005) 169.




\bibitem{cha}
A.~Linde, {\em Physics Letters B} {\bf 129}  (1983)   177.

\bibitem{lindeHybridInflation1994}
A.~Linde, {\em Phys. Rev. D} {\bf 49} (1994) 748.

%\bibitem{PhysRevD.95.083519}
%G.~A. Palma, B.~Pradenas, W.~Riquelme and S.~Sypsas, {\em Phys. Rev. D} {\bf 95} (2017)   083519.

\bibitem{PhysRevD.56.653}
S.~Khlebnikov and I.~Tkachev, {\em Phys. Rev. D} {\bf 56} (1997) 653.


\bibitem{PhysRevD.49.739}
A.~R. Liddle, {\em Phys. Rev. D} {\bf 49} (1994) 739.



%\bibitem{PhysRevD.83.083522}
%Z.-K. Guo, D.~J. Schwarz and Y.-Z. Zhang, %{\em Phys. Rev. D} {\bf 83} (2011)   083522.

\bibitem{aghanimPlanck2018Results2020}
N.~Aghanim and et~al.(Planck Collaboration), {\em Astronomy \& Astrophysics}
  {\bf 641} (2020).

%\bibitem{PhysRevD.47.426}
%F.~C. Adams, J.~R. Bond, K.~Freese, J.~A. %Frieman and A.~V. Olinto, {\em Phys.
 % Rev. D} {\bf 47} (1993) 426.

\bibitem{Boubekeur_2005}
L.~Boubekeur and D.~H. Lyth, {\em Journal of Cosmology and Astroparticle
  Physics} {\bf 07} (2005)   010.

\bibitem{Shamit}
S.~Kachru, R.~Kallosh, A.~Linde, J.~Maldacena, L.~McAllister and S.~P. Trivedi,
  {\em Journal of Cosmology and Astroparticle Physics} {\bf 10} (2003)
  013.

\bibitem{planckcollaborationPlanck2018Results2020}
Y.~Akrami and et~al.(Planck Collaboration), {\em Astronomy \& Astrophysics}
  {\bf 641}  (2020)   A10.

\bibitem{PhysRevLett.73.1886}
G.~Dvali, Q.~Shafi and R.~Schaefer, {\em Phys. Rev. Lett.} {\bf 73} (1994)
  1886.



\bibitem{Ahmed}
W.~Ahmed, A.~Karozas, G.~K. Leontaris and U.~Zubair, {\em Journal of Cosmology
  and Astroparticle Physics} {\bf 06} (2022)   027.

\bibitem{clarkeConstraintsPrimordialGravitational2020}
T.~J. Clarke, E.~J. Copeland and A.~Moss, {\em Journal of Cosmology and
  Astroparticle Physics} {\bf 10} (2020)   002.

\bibitem{garcia-bellidoStochasticBackgroundGravitational2007}
J.~Garc\'{\i}a-Bellido and D.~G. Figueroa, {\em Phys. Rev. Lett.} {\bf 98} (2007)   061302.

\bibitem{Choi_2021}
Ki-Young Choi, Su-beom Kang, Rathul Nath Raveendran, {\em Journal of Cosmology and
  Astroparticle Physics} {\bf 06} (2021)   054.

\bibitem{joyNewUniversalLocal2008}
M.~Joy, V.~Sahni and A.~A. Starobinsky, {\em Phys. Rev. D} {\bf 77} (2008)
   023514.

\bibitem{Minu Joy_2011}
M.~Joy, T.~Souradeep, {\em Journal of Cosmology and Astroparticle Physics} {\bf 02} (2011) 016

\bibitem{Mukherjee_2015}
S.~Mukherjee, S.~Das, M.~Joy, T.~ Souradeep, {\em Journal of Cosmology and Astroparticle Physics}  {\bf 01} (2015) 043

%\bibitem{Stein_2022}
%N.~K. Stein and W.~H. Kinney, {\em Journal %of Cosmology and Astroparticle
 % Physics} {\bf 01} (2022)   022.
 
\bibitem{liddleCosmologicalInflationLargeScale2000}
A.~R. Liddle and D.~H. Lyth, {\em Cosmological Inflation and Large-Scale
  Structure} (Cambridge University Press, 2000).


\bibitem{MinuJoy_2009}
M.~Joy, A. Shafieloo, V. Sahni and  A. A. Starobinsky, {\em Journal of Cosmology and Astroparticle Physics} {\bf 06} (2009)   028.
  
\bibitem{Gong_2022}
J.-O. Gong and M.~Mylova, {\em Journal of Cosmology and Astroparticle Physics}
  {\bf 07} (2022)   021.

\bibitem{bunnFouryearCOBENormalization1996}
E.~F. Bunn, A.~R. Liddle and M.~White, {\em Phys. Rev. D} {\bf 54} (1996)
  R5917.
\bibitem{PhysRevD.96.063508}
Sheere, Connor, Van Engelen et.al., {\em Phys. Rev. D} {\bf 96} (2017) 063508


%\bibitem{mis}
%Y.~Watanabe and E.~Komatsu, {\em Phys. Rev. D} {\bf 73} (2006)   %123515.

%\bibitem{G_2021}
%C.~Gómez and R.~Jimenez, {\em Journal of Cosmology and %Astroparticle Physics}
 % {\bf 10} (2021)   055.

\bibitem{keckarrayandbicep2collaborationsConstraintsPrimordialGravitational2018}
 Keck Array and BICEP2 Collaborations (P.~A.~R. Ade {\it et al}.), {\em Phys. Rev. Lett.} {\bf 121} (2018) 221301.




\bibitem{PhysRevD.73.123515}
Watanabe, Yuki and Komatsu, Eiichiro, {\em Phys. Rev. D} {\bf 73} (2006)   123515.

\bibitem{caprini}
Chiara Caprini, Daniel G Figueroa, {\em Class. Quantum Grav} {\bf 35}(2018) 163001.

\bibitem{Chowdhury_2022}
D.~Chowdhury, G.~Tasinato and I.~Zavala, {\em Journal of Cosmology and
  Astroparticle Physics} {\bf 08} (2022)   010.

\bibitem{PhysRevD.48.4613}
Turner, S.~Michael, White, Martin, Lidsey, E.~James, {\em Phys. Rev. D} {\bf 48}(1993) 4613.



\bibitem{laskyGravitationalWaveCosmology292016}
P.~D. Lasky{it et al}, {\em Phys. Rev. X} {\bf 6} (2016)
  011035.

\bibitem{Figueroa_2019}
D.~G. Figueroa and E.~H. Tanin, {\em Journal of Cosmology and Astroparticle
  Physics} {\bf 08} (2019)   011.

\bibitem{PhysRevD.78.043531}
L.~A. Boyle and A.~Buonanno, {\em Phys. Rev. D} {\bf 78} (2008)   043531.

\bibitem{campetiMeasuringSpectrumPrimordial2020}
P.~Campeti, E.~Komatsu, D.~Poletti and C.~Baccigalupi, {\em Journal of
  Cosmology and Astroparticle Physics} {\bf 01} (2021)   012.

\bibitem{Podolsky2005EquationOS}
D.~Podolsky, G.~N. Felder, L.~Kofman and M.~Peloso, {\em Phys. Rev. D} {\bf 73}
  (2006)   023501.

\bibitem{PhysRevLetDai}
L. Dai, M. Kamionkowski and J. Wang, {\em Phys. Rev. Lett.} {\bf 113} (2014) 041302.

\bibitem{PhysRevDJul}
J. B. Mu\~noz and M. Kamionkowski, {\em Phys. Rev. D} {\bf
  91} (2015)   043521.



\bibitem{PhysRevLett.73.3195}
L.~Kofman, A.~Linde and A.~A. Starobinsky, {\em Phys. Rev. Lett.} {\bf 73} (1994) 3195.





\bibitem{garcia-bellidoGravitationalWaveBackground2008}
J.~Garc\'{\i}a-Bellido, D.~G. Figueroa and A.~Sastre, {\em Phys. Rev. D} {\bf
  77} (2008)   043517.






\end{thebibliography}

\end{document}